\documentclass{article}
\usepackage{blindtext}
\usepackage{geometry}
\usepackage{seqsplit}
\usepackage{graphicx}
\usepackage{float}
\usepackage{gensymb}
\usepackage{breqn}
\usepackage{amsmath}
\usepackage{textcomp}
\usepackage{color}
\usepackage{tabularx}
\usepackage{doi}
\usepackage{caption}
\usepackage{subcaption}
\usepackage{doi}
\usepackage{url}
\usepackage[
    backend=biber,
    style=numeric,
    natbib=true,
    url=false, 
    doi=true,
    eprint=false, 
    sorting=none
]{biblatex}
\usepackage{hyperref}

\makeatletter
\let\@fnsymbol\@arabic
\makeatother

\title{Drone flight data reveal energy and greenhouse gas emissions savings for small package delivery}
\author{Thiago A. Rodrigues\thanks{Department of Civil and Environmental Engineering. Carnegie Mellon University. 5000 Forbes Avenue, Pittsburgh, 15213, PA USA} \textsuperscript{,}\thanks{Corresponding authors: {\tt\small{tarodrig@andrew.cmu.edu}} and {\tt\small{csamaras@cmu.edu}}}, \;Jay Patrikar\thanks{Robotics Institute, Carnegie Mellon University, 5000 Forbes Avenue, Pittsburgh, 15213, USA}, \;Natalia L. Oliveira\thanks{Department of Statistics and Data Science, Carnegie Mellon University, 5000 Forbes Avenue, Pittsburgh, 15213, USA} \textsuperscript{,}\thanks{Machine Learning Department, Carnegie Mellon University, 5000 Forbes Avenue, Pittsburgh, 15213, USA}, \\H. Scott Matthews\footnotemark[1], \;Sebastian Scherer\footnotemark[3], \;Constantine Samaras \footnotemark[1] \textsuperscript{,} \footnotemark[2]
}

\date{}
\begin{document}

\maketitle

\begin{abstract}
The adoption of Uncrewed Aerial Vehicles (UAVs) for last-mile deliveries will affect the energy productivity of package delivery and require new methods to understand the associated energy consumption and greenhouse gas (GHG) emissions. Here we combine empirical testing of 187 quadcopter flights with first principles analysis to develop a usable energy model for drone package delivery. We develop a machine learning algorithm to assess energy use across three different flight regimes: takeoff, cruise, and landing. Our model shows that, in the US, a small electric quadcopter drone with a payload of 1 kg would consume approximately 0.05 MJ/km and result in 41 g of CO$_{2}$e per package. The energy per package delivered by drones (0.19 MJ/package) can be up to 96\% lower than conventional transportation modes. Our open model and generalizable coefficients can assist stakeholders in understanding and improving the energy use of drone package delivery.
\end{abstract}

Key words: quad-copter drone, last-mile delivery, energy consumption, greenhouse gas emissions, robot delivery, autonomous delivery.

\section{Introduction}

Achieving large improvements in the energy productivity of the freight transportation sector is challenging, especially in the overwhelmingly petroleum-powered transport sector where medium and heavy trucks in the US comprises 24\% of transportation energy use. This sector is responsible for 37\% of transportation-related greenhouse gas (GHG) emissions while light-duty vehicles comprise 57\% of transportation GHG emissions and 64\% of transportation energy use. In addition, transportation remains a large source of nitrogen oxides (NOx) and other air pollutants \cite{Davis2021Transportation39}. However, the way that consumers are obtaining goods in the U.S. is changing rapidly \cite{Dost2018}.

Even before COVID-19, the growing demand for fast, contactless deliveries has been driving firms to experiment with automated package delivery vehicles, such as Uncrewed Aerial Vehicles (UAVs) that can avoid traffic in urban centers and rapidly reach rural areas that would not be served otherwise \cite{Mims2020, Lee2020TheFuture}. Initial survey data of 483 customers in Portland, Oregon by Pani et al. \cite{Pani2020EvaluatingPandemic} show that COVID-19 is contributing to an environment where more than 60\% of online customers are willing to pay extra to receive their packages using autonomous delivery robots. Nevertheless, along with technology and policy challenges, increasing shipping costs is one of the main limitation for the adoption of autonomous delivery vehicles \cite{Kapser2020AcceptancePerceptions}.

The appeal of delivery robots also reflects new physical distancing demands to avoid the spread of coronavirus in product deliveries \cite{Abrar2020AnSystem}. As autonomous delivery technologies advance, new companies emerge to compete for this market niche \cite{Joerss2016HowMcKinsey, Bensinger2016AmazonsFedEx, Gross2013, Murphy2016, UPS2017}. At the same time, alternative transport modes, such as electric cargo bicycles, are becoming cost-effective alternatives to delivery trucks for short-distance deliveries \cite{Sheth2019MeasuringAreas}, drastically reducing the $CO_2$ emissions of last-mile delivery in highly dense metropolitan areas \cite{McLeod2020QuantifyingDelivery}. With the increased electrification of delivery vehicles, the energy consumption and environmental impacts of the transportation sector are expected to change drastically over the coming years \cite{Gebeloff2017, Dominkovic2018TheTransition}, and both technology and demand are primary drivers. Widespread adoption of UAVs to replace a portion of first/last-mile truck pickups and deliveries could reshape this sector by changing demand patterns and shifting fuel demands from fossil fuels to electricity. Autonomous Delivery Robots are coming to the transportation sector, but how these vehicles and systems could be designed to maximize energy productivity is less clear.

So far, a few studies have estimated the energy consumption of quadcopter vehicles and the energy estimations vary considerably among the different methods used \cite{Zhang2021EnergyAssessment}. Some studies have created models based on theoretical principles \cite{DAndrea2014GuestDeliver, Figliozzi2017LifecycleEmissions, Dorling2016VehicleDelivery, Liu2017, Lohn2018TheBuzz, Stolaroff2018EnergyDelivery, Jeong2019Truck-droneZones,Torabbeigi2019, Kirschstein2020ComparisonServices}, while others have developed models based on regression models built on small flight samples \cite{Dorling2016VehicleDelivery, Tseng2017AutonomousDrones, Tseng2017FlightDrones, Jeong2019Truck-droneZones}. Finally, a comparison of the energy consumption and greenhouse gas (GHG) emissions between package delivery UAVs and different transportation modes have been estimated by a few studies \cite{Stolaroff2018EnergyDelivery, Figliozzi2020CarbonVehicles, Goodchild2018DeliveryIndustry, Borlaug2021Heavy-dutySystems}, but alternative emerging delivery modes, such as electric cargo bicycles are not included.

Here we help stakeholders and researchers understand the energy use of uncrewed aerial package delivery drones. We provide an energy model based on extensive empirical data from 187 flights of a quadcopter drone DJI Matrice 100, from which we developed a novel and high-resolution dataset \cite{Rodrigues2020a} of package delivery drone energy use. We also develop an algorithm that automatically identifies the flight regime across takeoff, cruise and landing. We show the impact of the cruise speed and payload mass on the drone's range, provide generalizable energy use coefficients, and we compare the energy consumption and GHG emissions of a small quadcotper drone to other last-mile transportation modes on a energy per package basis.

\section{Results and Discussion}

We collected data on 187 flights to assess the power profile of a package delivery drone given a set of operational parameters (payload, altitude, and speed during cruise). The data, available at \cite{Rodrigues2020a}, and a data descriptor \cite{Rodrigues2021In-flightDelivery}, provide the details of the experiment. In addition, we have developed an algorithm that separates the data into three different flight regimes: takeoff, cruise, and landing, in order to better understand the energy consumption profile during flight, see Supplementary Figures S1 to S4. 

Then, we conducted a first principles analysis and developed a model to estimate the energy required to power a quadcopter. Each of the flight regimes were modeled separately, that is, each energy model was treated as a model class and three different optimal models from that class were selected, one per regime. In order to fairly compare the model classes' performance and avoid overfitting, we split the data into train and test folds following a stratification strategy by flight ID number. With 120 flights, the training fold was used to estimate the parameters of each model, which were then applied to the remaining 67 flights from the test fold in order to evaluate the performance of the energy models on unseen data.  

\subsection{Energy Model}
Our energy (E) model uses the Induced Power ($P_{i}$), which is the power required to overcome gravity in a hover-no-wind situation, as a parameter estimator of the average power observed throughout the flight.

\begin{equation}
    E = (b_{1}P_{i} + b_{0})t
    \label{eq:linearModel}
\end{equation}
where, $t$ is the flight duration, and $b_{1}$ and $b_{0}$ are coefficients that linearly correlate $P_{i}$ and the average power throughout the flight.
The induced power, used in eq. \ref{eq:linearModel}, is calculated as 
\begin{equation}
    P_{i} = \frac{\left(mg\right)^{1.5}}{\sqrt{2\rho A}}
\end{equation}
where, $m$ is the total mass of the drone (including the payload), $g$ is the acceleration of gravity, $\rho$ is the air density, and $A$ is the total area under the propellers. 

The estimated coefficients and their standard errors are shown in Table \ref{Tab:coefficients_Method1}. 

\begin{table}[H]
\centering
\caption{Estimated coefficient $\pm$ bootstrap standard error}
\begin{tabular}{llll}
\textbf{Coef.} & \textbf{Take off} & \textbf{Cruise} & \textbf{Landing} \\ \hline
$b_1$ &	1.97 $\pm$ 0.08 &	1.69 $\pm$ 0.06 &	1.62 $\pm$ 0.14\\
$b_0$ &	13.8 $\pm$ 0.01 &	16.8 $\pm$ 0.01 &	-4.7 $\pm$ 0.01\\  \hline
\end{tabular}
\label{Tab:coefficients_Method1}
\end{table}

The coefficients shown in Table \ref{Tab:coefficients_Method1} were obtained by performing a linear regression between $P_{i}$ and the average power observed throughout each of the 120 flights. The results were then applied to the remaining flights and the absolute relative error was 3\% on average, proving the accuracy of the energy model in terms of estimation of energy consumption.

\begin{figure}[H]
    \centering
    \includegraphics[width=\textwidth]{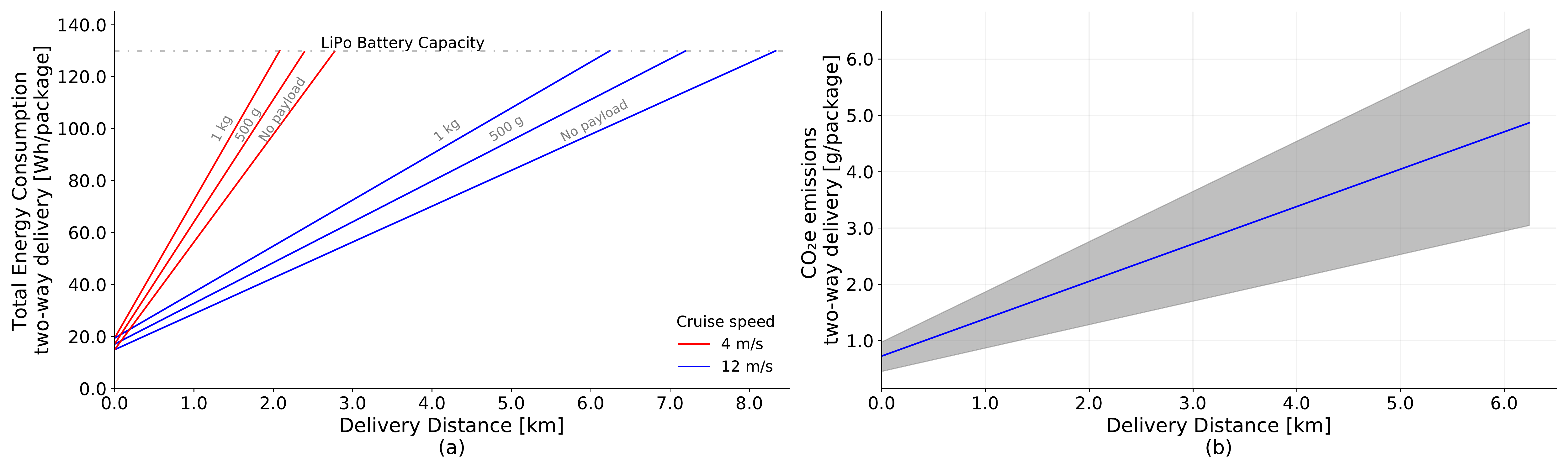}
    \caption{(a) Total energy consumption by distance of delivery varying payload mass and cruise speed. (b) CO$_{2}$e emissions according to the delivery distance varying according grid emission factor and battery life cycle emissions.  The total energy and CO$_{2}$e correspond to takeoff, cruise from the origin to destination and landing loaded, and takeoff, cruise from destination to origin and landing empty. As an energy limitation, the nominal capacity of LiPO TB48D battery is 130 Wh \cite{DJI}. Altitude during cruise of 100 m, takeoff speed 2.5 m/s and landing speed 2 m/s.\label{fig:dist}}
\end{figure}

With the energy model validated, we estimated the energy consumption of a  package delivered by a small quadcopter drone. Figure \ref{fig:dist}a shows the total Energy Consumption for a two-way delivery trip (delivery and return) according to the delivery distance and the total weight of the drone (with payload) operating at a cruise speed of 4 and 12 m/s and altitude of 100 m. Our analysis also shows that variations in the cruise speed have great impact on the total energy consumption per trip and consequently range of the drone (Figure \ref{fig:dist}a). The total time of flight is reduced as the speed increases, which results in longer distances for the same amount of energy. 
In addition, we calculated the GHG emissions per package delivered based on the US electricity grid, upstream electricity generation, and battery life-cycle emissions according to the delivery distance (Figure \ref{fig:dist}b)

\subsection{Comparison between different transportation modes}
We compared the energy consumption of quadcopter drones against diesel and electric medium-duty trucks and small vans, and electric cargo bicycles. 
   
\begin{figure}[H]
    \centering
    \includegraphics[width=\textwidth]{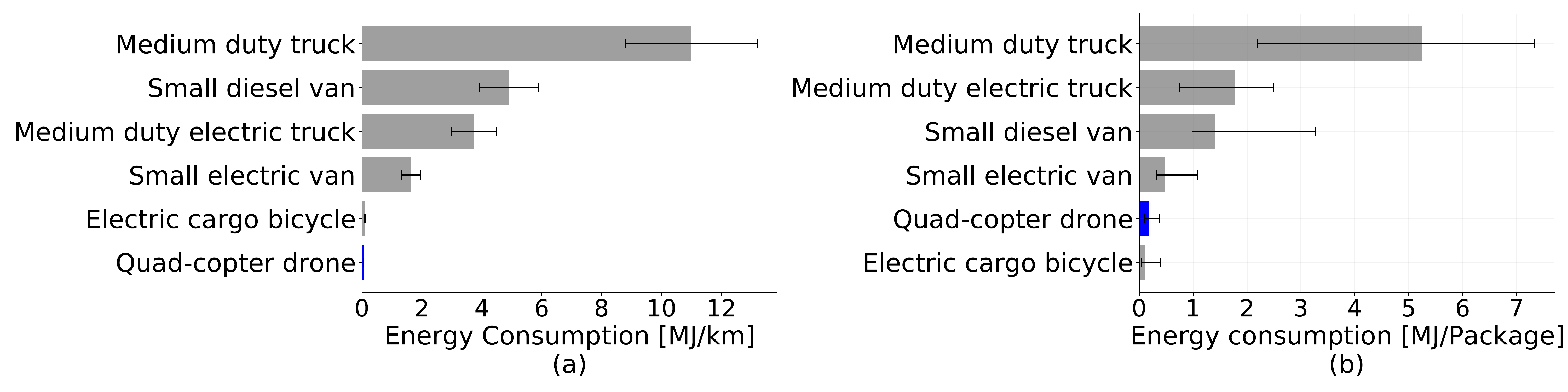}
    \caption{Energy Consumption for different transportation modes. Error bars represent variations in (a) driving styles and vehicle characteristics, and (b) number of packages delivered per distance.}
    \label{fig:f2}
\end{figure}   
   
Total energy consumption per distance of small quadcopter drones is among the lowest across transportation modes, as the vehicle is small, light, and has lower payload capacity (Figure \ref{fig:f2}a). However Figure \ref{fig:f2}b shows the energy consumption per package of drone-equivalent deliveries, i.e., assuming that all packages delivered by the other modes are within the payload and space capacity of a small quadcopter drone \cite{Guglielmo2013TurnsPounds}. On an energy consumption per package basis, small quadcopter drones are also among the most efficient methods of delivery. The number of stops per kilometer and the number of packages delivered per stop varies according to the transportation mode and delivery density (highly dense areas are more likely to have more stops and packages delivered per kilometer).         

\begin{figure}[H]
    \centering
    \includegraphics[width=\textwidth]{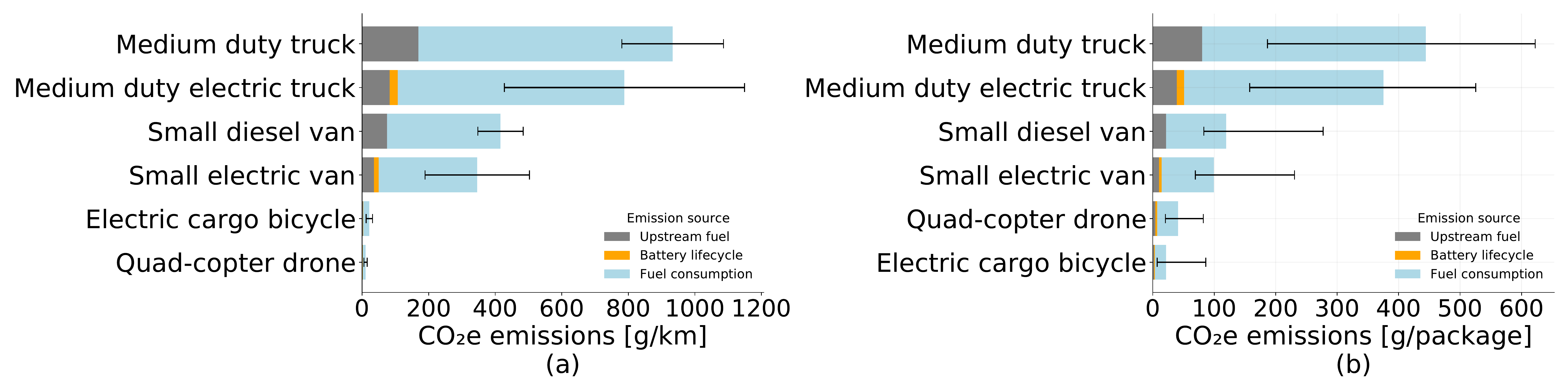}
    \caption{Greenhouse gas (GHG) emissions for different transportation modes. Error bars represent uncertainties due to variations on fuel carbon intensity, battery life cycle emissions, and number of packages delivered per distance.}
    \label{fig:f3}
\end{figure}

Similarly, an analysis of the greenhouse gas (GHG) emissions of the fuel of each transportation mode shows that quadcopter drones are     among the most efficient vehicles in grams of CO$_{2}$e per km (Figure \ref{fig:f3}a) and a competitive alternative in terms of GHG emissions per package (Figure \ref{fig:f3}b). On the other hand, it is important to note that small drones are considerably limited in terms of weight and volume of the packages transported. Therefore, an analysis of the energy consumption and GHG emissions on a per metric ton-km basis (Supplementary Figure S5) shows that small drones are the most energy-intensive vehicles. Also, local airspace regulations that require longer delivery routes, not considered in this study, can potentially increase  the energy consumption and GHG emissions of drones \cite{Elsayed2020TheOperation}. Finally, alternative methods, such as the concept of mobile warehouses \cite{SrivatsaSrinivas2021MovingBeyond} can be an effective alternative to incorporate drones and mitigate current limitations by combining drones and delivery trucks.   

We compared our results to values provided by the United Parcel Service, Inc. (UPS). In 2019, UPS  reported the energy intensity for U.S. Domestic Package operations was 28 MJ/package, from which ground vehicles represented approximately 9.5 MJ/Package or 34\% (airline fuel, facility heating fuel and indirect energy correspond to 60\%, 3\% and 3\%, respectively), with GHG emissions (CO$_{2}$e) intensity of 1 kg/package \cite{UPS20192019Index}. It is important to note that these values encompass the entire ground fleet, rather than only last mile delivery.

We estimate a small quadcopter drone, with a payload of 1 kg operating at a cruise speed of 12 m/s and cruise altitude of 100 m, consumes approximately 0.05 MJ/km and is generates 41 g of CO$_{2}$e per package when charged on average U.S. electricity.  Our energy model has simple, generalizable, and accurate coefficients that can provide stakeholders and researchers an energy consumption estimation for speed ranges below 12 m/s. However, at greater speeds or using drones with more surface area, a more comprehensive energy profile method could provide more accurate predictions. 

The energy consumption of small quadcopter drones is comparable to the most energy efficient modes of last-delivery when the total mass of delivery is not the main feature considered. For example, in delivery situations where small and light items with high added value, such as small electronics and medicines, drones might became a competitive tool to reduce transportation emissions in large urban centers \cite{Isik2021TransportationRates}. In these scenarios, we found that drones can reduce the energy consumption by 96\% and 60\% and GHG emissions by 91\% and 59\% per package delivered by replacing diesel trucks and electric vans, respectively. We also found that the delivery intensity, i.e. the number of packages delivered per km, and the fuel carbon intensity are the main factors contributing to the drone's energy and environmental performances. It is also important to note that the drone used to collect the data was not optimized to minimize energy consumption, which could further improve the its efficiency.  

\section{Methods}

\subsection{Experiment}
We performed a series of flights to empirically measure the energy consumption of a quadcopter UAV. An experimental protocol was created and followed to ensure a reliable approach for data acquisition \cite{Rodrigues2021In-flightDelivery}. 

A DJI\textsuperscript{\textregistered}  Matrice 100 (M100) quadcopter was equipped with an anemometer, current and voltage monitor, GPS, and accelerometer collecting data on wind speed and direction, battery current and voltage demand, and position, orientation, velocity and acceleration. The flights were performed in a pre-established route with varying altitude (25 m, 50 m, 75 m and 100 m), speed (4 m/s, 6 m/s, 8 m/s, 10 m/s and 12 m/s) and payload mass (no payload, 250g and 500g). Each combination was repeated at least three times, totaling 187 flights. The data provided by each sensor were synchronized to a frequency of approximately 5Hz using the ApproximateTime \cite{Foundation2010} message filter policy of Robot Operating System (ROS).

For a better understanding of the energy consumption profile of each flight we created an algorithm to automatically divide the data into three different flight regimes: takeoff, cruise, and landing (see Supplementary Figures S1 to S4).

\subsection{First Principles Analysis}
The energy required to power a UAV can be estimated using a first principle analysis based on helicopter aerodynamics \cite{Rotaru2017a}. First, we defined the working coordinate frames for a quadcopter drone (see Supplementary Figures S6 to S8). Then, the we assessed the power required to maintain the drone at a steady hover condition. Finally, we expanded the power analysis to include other power demands.

\begin{figure}[H]
    \centering
    \includegraphics[width = 8 cm]{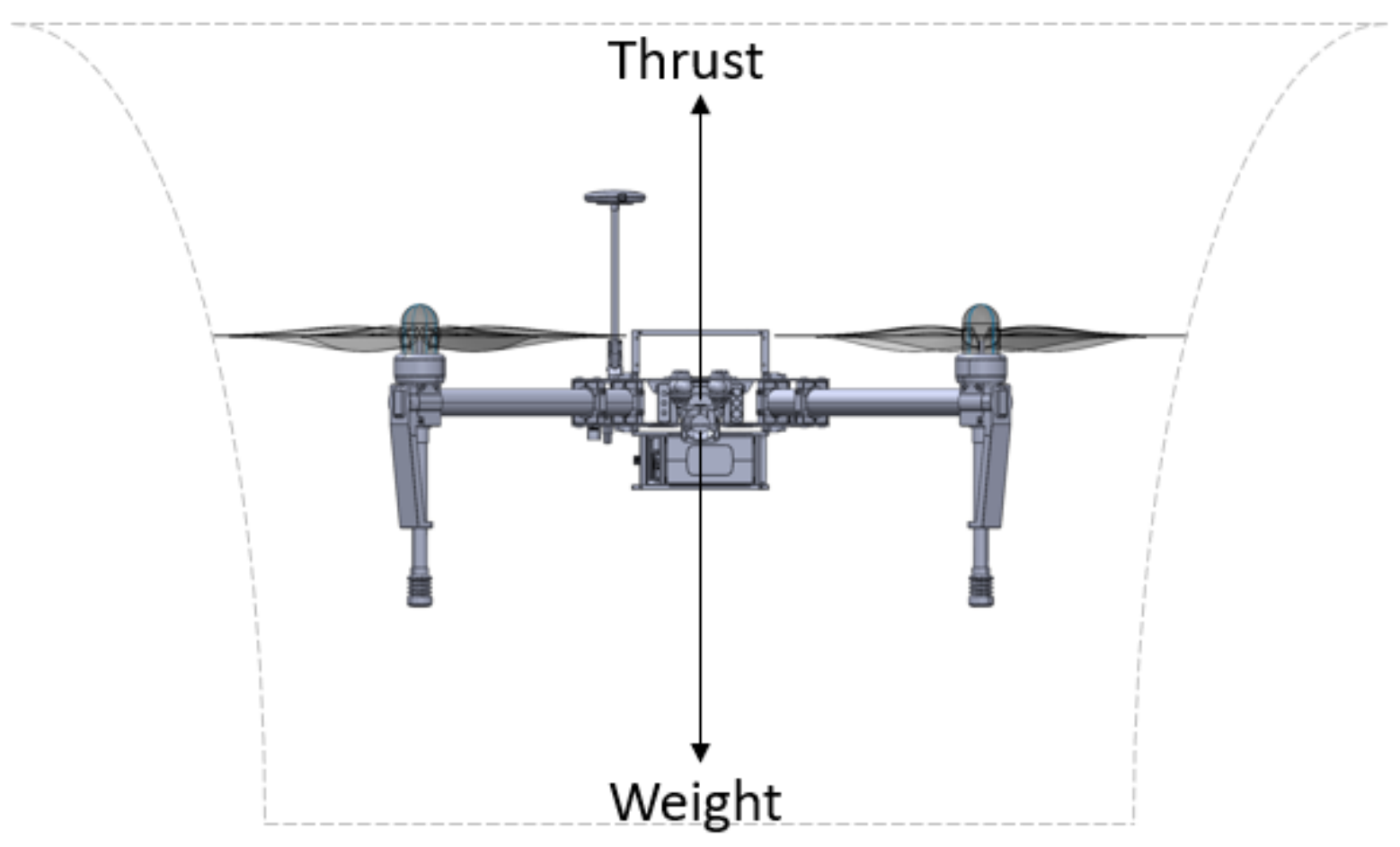}
    \caption{Hover condition without wind.}
    \label{fig:hover}
\end{figure}

The main power demand of a drone is in the form of induced power ($P_{i}$). The induced power represents the power required to overcome the force of gravity in order to keep the aircraft in the air, and it can vary according to the flight maneuver \cite{Rotaru2017a}. The most basic way to estimate $P_{i}$ is considering a hover condition without wind (Figure \ref{fig:hover}). 

In that case, the thrust ($T$) equals the only force acting on the drone, its weight ($W = mg$) \cite{Rotaru2017a}, and $P_{i}$ can be estimated as
\begin{equation}
    P_{i} = T v_{i}
    \label{eqn:Pi}
\end{equation}
where $v_{i}$ is the induced velocity.

During hover, $v_{i}$ can be simplified as
\begin{equation}
    v_{i} = \sqrt{\frac{T}{2 \rho A}}
    \label{eqn:vi_hover}
\end{equation}
where $\rho$ is the air density and $A$ is the total area covered by the all four propellers. 

Combining Eq. \ref{eqn:vi_hover} and \ref{eqn:Pi}
\begin{equation}
    P_{i} = \frac{(T)^{3/2}}{\sqrt{2 \rho A}} =\frac{(m g)^{1.5}}{\sqrt{2 \rho A}}
    \label{pi_hover}
\end{equation}
where $m$ is the total mass of the drone, $g$ is the gravitational acceleration.  

More details of the first principles analysis and an expanded first-principles energy model is available in the supplementary information (see supplementary Figures S9 to S11).  

\subsection{Energy Model}
Our energy model inquires how effectively $P_{i}$ can be used as an estimator for the energy consumed during a package delivery flight. In such a case, the average power ($\Bar{P}$)
throughout the flight is modeled as a linear regression of the induced power 
\begin{equation}
    \Bar{P} = b_{1}P_{i} + b_{0}
    \label{eqn:E_1}
\end{equation}
where $b_{1}$ and $b_{0}$ are the slope and intercept of the linear regression, respectively.  

\begin{figure}[H]
    \centering
    \includegraphics[width = 8 cm]{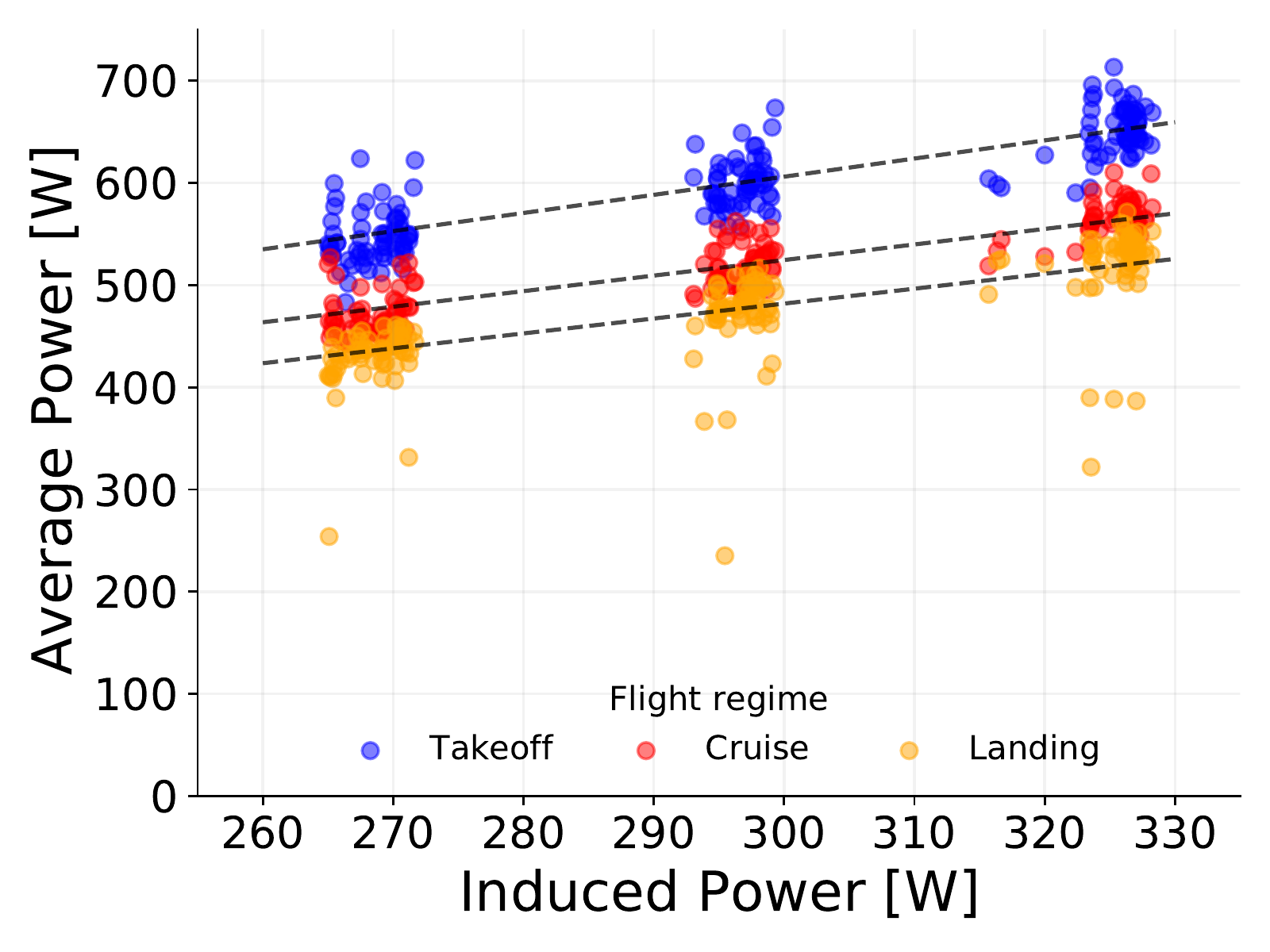}
    \caption{Linear regression of flights separated by flight regime.}
    \label{fig:LN}
\end{figure}

Eq. \ref{eqn:E_1} is expanded to account for the sum of the three flight regimes (Figure \ref{fig:LN}) and the total energy consumption ($E$) is estimated as
\begin{equation}
    E = \sum_{r\in\mathcal{R},\; l\in\mathcal{L}}(b_{1}^{(r,\;l)}P_{i}^{(l)} + b_{0}^{(r,\;l)})t^{(r,\;l)}
    \label{eqn:E_sum}
\end{equation}
for $\mathcal{R} = \{takeoff,cruise,landing\}$ and $\mathcal{L} = \{loaded, unloaded\}$. 

\subsection{Machine Learning Approach} 
However, evaluating if the model's performance is good given the available measurements cannot be inferred from its performance alone. Therefore, we compare the predictive power of the energy model to a flexible nonlinear algorithm \cite{xgboostRpackage}, \texttt{XGBoost}, available in the programming environment \texttt{R}. This boosted tree algorithm prioritizes predictive power against interpretability, and it is appropriate for predictive performance given the available features. If our energy model presents similar accuracy to \texttt{XGBoost}, it indicates that the parametric and functional restrictions we have made for the energy model development are suitable. 

We fitted a gradient boosted tree algorithm, \texttt{XBGoost} \cite{xgboostRpackage}. The algorithm was separately trained for each flight regime with a quadratic loss function and for all regimes, we used 75\% as a subsample ratio of both features and observations for each tree.  For hyperparameter tuning, we varied learning rate, maximum tree depth, and regularization parameter $\gamma$ in a grid search approach. 
5-fold CV was used for error estimation; for tuning only,  we compared Absolute Relative Error (ARE, Equation \ref{eq:relativeerror}) instead of quadratic error. After tuning, the model was trained with the optimal hyperparameters on the entire training set and AREs were computed for the flights on the test set. 

\paragraph{Standard errors} 
To obtain standard errors of the estimated coefficients, we used a nonparametric bootstrap approach \cite{EfroTibs93}. 1000 bootstrap replications were used to resample with replacement the 120 training flights, and the two energy models for the three flight regimes were refitted for each bootstrap sample. At the end, standard error of the coefficients were obtained from their sampling distribution. 

\paragraph{Model predictive power} 
To evaluate the model's predictive power, the regime-specific fitted models were then applied to the testing flights of the test set and were compared by Absolute Relative Error (ARE), computed at flight resolution. That is, for each flight from the test set, we computed their $E_{measured}$ integrating power over time, and the $E_{estimated}^m$ as the sum of the integral of the estimated power over time for the three flight regimes via method $m$: 
\begin{equation}
\label{eq:relativeerror}
    ARE(m) = \left | \frac{E_{measured}-E_{estimated}^m}{E_{measured}} \right|
\end{equation}
for $m \in \{\textit{Energy Model}, \textit{XBGoost}\}. $

\subsection{Drone's range}
Our analysis also shows the impact of varying operational parameters (speed, altitude and payload) on the range of the drone. The two-way drone range (d) can be calculated considering the cruise speed
\begin{equation}
    E = \sum(b_{1}P_{i} + b_{0})t = \sum E_{\,takeoff} + \sum E_{\,landing} + \sum (b_{1}P_{i} + b_{0}) \frac{d}{V_{cr}}
\end{equation}
Expanding for a two-way trip and solving for d

\begin{equation}
    d = \frac{
    \left[E_{max} - \left(E_{\,takeoff}^{\,l} + E_{\,takeoff}^{\,u} + E_{\,landing}^{\,l} + E_{\,landing}^{\,u} \right)\right]V_{cr}}{b_{1} \left(P_{i}^{\,l}+P_{i}^{\,u}\right)  + 2b_{0} }
\end{equation}
where, $E_{max}$ is the energy available in the battery; $E_{takeoff}$ and $E_{landing}$ are the energy consumed during takeoff and landing for delivery (loaded = $l$) and returning (unloaded = $u$), respectively;  $P_{i}$ is the induced power, calculated using Eq. \ref{pi_hover}, for delivery (loaded = $l$) and return (unloaded = $u$); $b_{1}$ and $b_{0}$ the coefficients from Table \ref{Tab:coefficients_Method1} for cruise and $V_{cr}$ is the average inertial cruise speed.

The energy during takeoff and landing can be calculated as
\begin{equation}
    E = \left(b_{1}P_{i} + b_{0}\right)\frac{h}{V}
\end{equation}
where h is the cruise altitude, V is the average speed during takeoff and landing. 

For instance, a small quadcopter operating at $V_{cr}$ = 12 m/s, payload = 1000 g, h = 100 m, takeoff average speed ($V_{tk}$) = 2.5 m/s, landing average speed $V_{ld}$ = 2.0 m/s has a range of approximately 11 km (5.7 km of delivery range), consuming approximately 120 Wh (per round trip delivery). 
Therefore, a quadcopter drone flying under these conditions would consume approximately 0.039 MJ/km, not considering charging and transmission losses. The energy consumption during takeoff and landing for this trip corresponds to approximately $17\%$ of the total energy consumption (19.4 Wh per trip). This share of energy could be reduced by 95\% in a 5-m takeoff (from 19.4 to Y 0.97 Wh per trip), which could be achieved, for instance, if the drone would depart from the top of a building. This would reduce total trip energy by 15\% (120 to 102 Wh), or increase the range from 5.7 to 6.7 km.  

\subsection{Transport mode comparison}
We compare the small quadcopter drone to different transportation modes in terms of energy consumption and CO$_{2}$e emissions and validate it against top-down sustainability reports from the United Parcel Service, Inc. (UPS)'s 2019 . The energy consumption of a medium-duty diesel truck is considered as 11 MJ/km \cite{Prohaska2016FieldTrucks}. Whereas a Medium-duty electric truck has energy consumption of 1.4 kWh/mile\cite{Prohaska2016FieldTrucks}, or 3.13 MJ/km. Diesel Vans operate at 18.4 MPG on average\cite{AdvantageOutfitters2019BestMore}, or 4.9 MJ/km, (conversion factors: 1 gallon of diesel = 137,381 Btu\cite{U.S.EnergyInformationAdministrationEIA2020EnergyExplained}, 1 MJ = 947.817 Btu, 1 mile = 1.60934 km). On the other hand, electric vans operate with energy consumption of 0.38 kWh/km\cite{Fiori2018ModellingDataset}, or 1.36 MJ/km. Finally, an electric cargo bicycle operates at 0.023 kWh/km\cite{Erlandsson2017TheVan}, or 0.08 MJ/km (conversion factors: 1 kWh = 3.6 MJ). In addition, variations in driving style can vary energy consumption by 40\%\cite{FuelEconomyManyMPG}. 
Based on our energy model, a small quadcopter drone consumes approximately 120 Wh in a 5.5 km delivery distance (11 km total distance), or 0.039 MJ/km, when delivering at maximum capacity (1 kg payload with unloaded return) and cruise speed of 12 m/s. Transmission losses of 6.5\% and a charging efficiency of 88\% \cite{ArgonneGREETPublication2016Cradle-to-GraveTechnologies,Shiau2009ImpactVehicles, USEPA2019EmissionseGRID, U.S.EnergyInformationAdministrationEIA2021HowStates} were included to the energy consumption of the electric vehicles (Table \ref{tab:energy_ghg_pack}). Supplementary Table S1 summarizes the nominal energy consumption and also provides the payload capacity of each mode. 

The electricity CO$_{2}$e emissions were considered to be the 2019 American average of 182 g/MJ (656 g/kWh), with the lower of 107 g/MJ (384 g/kWh) from New England and the upper limit of 249 g/MJ (896 g/kWh) reflecting non-baseload emissions from the central Mid-West\cite{USEPA2019EmissionseGRID}.
CO$_{2}$e emissions for Diesel Fuel combustion was considered as $1.61x10^{-4}$ lb/Btu\cite{U.S.EnergyInformationAdministrationEIA2020HowBurned}, or 69.35 g/MJ. Upstream GHG emissions for diesel and electricity generation are 15 g/MJ and 22 g/MJ\cite{Laboratory2020GREET}, respectively. The drone’s LiPO Battery life cycle emissions were assumed to be similar to Li-iron phosphate 2 g/MJ (base case), 0.6  g/MJ (low case) and 4 g/MJ (high case)\cite{Laboratory2020GREET}. Similarly, the electric cargo bicycle has battery life cycle emissions of 5.1 g/MJ (base case), 1.1  g/MJ (low case) and 16.9 g/MJ (high case), for the electric van 18.7 g/MJ (base case), 5.6  g/MJ (low case) and 37.4 g/MJ (high case), and for the medium duty electric truck  32.5 g/MJ (base case), 9.7  g/MJ (low case) and 65 g/MJ (high case).

The battery life cycle emissions for the drone (assumed to be similar to Li-iron phosphate) was calculated as 0.76 g/km (base case), 0.23  g/MJ (low case) and 1.52 g/MJ (high case). Similarly, the electric cargo bicycle has battery life cycle emissions of 1.3 g/km, considering a Li-ion NMC811 battery. For the electric van and electric medium duty truck we assumed a battery of Li-ion NMC811, resulting in 14.1 g/km for the van and 24.5 g/km for the truck.\cite{Laboratory2020GREET}

The energy consumption per package ($E_{pack}$) was calculated as
\begin{equation}
    E_{pack} = \frac{E_{dist}}{S_{freq}\cdot P_{freq}}
\end{equation}
where $E_{dist}$ is the energy consumption per distance unit, $S_{freq}$ is the number of stops to delivery packages per distance unit, and $P_{freq}$ is number of packages delivered per stop on average. 

Similary, the greenhouse gas emissions per package ($GHG_{package}$) is calculated as 

\begin{equation}
    GHG_{pack} = \frac{E_{dist}\cdot GHG_{energy}}{S_{freq}\cdot P_{freq}}
\end{equation}
where $GHG_{energy}$ is the mass of CO$_{2}$e per energy unit.

Table \ref{tab:energy_ghg_pack} summarizes the values calculated per vehicle.

% \usepackage{graphicx}
\begin{table}[H]
\centering
\caption{Base-case energy consumption and GHG emissions for different vehicles.}
\label{tab:energy_ghg_pack}
\resizebox{\textwidth}{!}{%
\begin{tabular}{lcccccc}
\textbf{Vehicle Class} &
  \multicolumn{1}{l}{\textbf{\begin{tabular}[c]{@{}l@{}}Energy Consumption\\ {[}MJ/km{]}\end{tabular}}} &
  \multicolumn{1}{l}{\textbf{\begin{tabular}[c]{@{}l@{}}Fuel GHG\\ emissions {[}g/km{]}\end{tabular}}} &
  \multicolumn{1}{l}{\textbf{\begin{tabular}[c]{@{}l@{}}Upstream GHG\\ emissions {[}g/km{]}\end{tabular}}} &
  \multicolumn{1}{l}{\textbf{\begin{tabular}[c]{@{}l@{}}Battery GHG\\ emissions {[}g/km{]}\end{tabular}}} &
  \multicolumn{1}{l}{\textbf{\begin{tabular}[c]{@{}l@{}}Energy consumption\\ {[}MJ/package{]}\end{tabular}}} &
  \multicolumn{1}{l}{\textbf{\begin{tabular}[c]{@{}l@{}}GHG emission\\ {[}g/package{]}\end{tabular}}} \\ \hline
Medium duty truck          & 11.00 & 764.5 & 168.7 &      & 5.24 & 444.4 \\
Small diesel van           & 4.90  & 340.6 & 75.2  &      & 1.41 & 119.5 \\
Medium duty electric truck & 3.74  & 681.4 & 82.4  & 24.5 & 1.78 & 375.4 \\
Small electric van         & 1.63  & 296.1 & 35.8  & 14.1 & 0.47 & 99.4  \\
Electric cargo bicycle     & 0.10  & 18.1  & 2.2   & 1.3  & 0.10 & 21.6  \\
Quad-copter drone          & 0.05  & 8.5   & 1.0   & 0.8  & 0.19 & 41.1  \\ \hline
\end{tabular}%
}
\end{table}

\section*{Acknowledgments}
This work was supported by the U.S. Department of Energy's Vehicle Technologies Office, Award Number DE-EE0008463. This article was prepared while C.S. was employed at Carnegie Mellon University, and is currently on public service leave. The opinions expressed in this article are the authors’ own and do not reflect the view of the United States government or any other organization.

\section*{Data Availability}
All drone data is available is at \href{https://doi.org/10.1184/R1/12683453}{https://doi.org/10.1184/R1/12683453} and the all the modeling code is available at \href{https://github.com/thiago-a-rod/energy\_consumption}{https://github.com/thiago-a-rod/energy\_consumption} 

\section*{Author Contributions Statement}
The research project was conceived by C.S, H.M. and S.S. The drone data collection experiment was designed by J.P. The energy model was developed by T.R. The results were analyzed by T.R. and N.O. All authors contributed to the writing of the manuscript. 

\section*{Declaration of Interests}
The authors declare no competing interests.

\section*{Supplementary Information}
Supplementary Information is available with this manuscript.

%\bibliographystyle{unsrt}
%\bibliography{references.bib}
\printbibliography

\end{document}